\documentstyle[11pt]{article}

\begin{document}

\renewcommand{\thefootnote}{\fnsymbol{footnote}}

{\large\centerline{\bf 
Glueball Masses from Hamiltonian Lattice \rm{QCD}
}

\vskip 1.0cm

\centerline{Lian HU$^{a,d}$, Xiang-Qian LUO$^{a,b,c}$, 
Qi-Zhou CHEN$^{a,b}$, Xi-Yan FANG$^{a,b,c}$, 
and Shuo-Hong GUO$^b$}

\vskip 1.0cm

\centerline{\it $^a$ CCAST (World Laboratory), 
Beijing 100080, China}

\centerline{\it $^b$ Department of Physics, 
Zhongshan University, Guangzhou 510275, China 
\footnote{Mailing address}}

\centerline{\it $^c$ Center for Computational Physics, 
Zhongshan University, Guangzhou 510275, China}

\centerline{\it 
$^d$Department of Physics, South China Normal University, 
Guangzhou 510630, 
China
}

\vskip 3.0cm

\begin{abstract}
We calculate the masses of the $0^{++}$, $0^{--}$ and $1^{+-}$ 
glueballs from \rm{QCD} in 3+1 dimensions using 
an eigenvalue equation method for Hamiltonian lattice QCD 
developed and described in \cite{GCL,QCD3,MASS} by the authors. 
The mass ratios become approximately constants
in the coupling region $6/g^2 \in [6.0,6.4]$, from which
we estimate $M(0^{--})/M(0^{++})=2.44 \pm 0.05 \pm 0.20$ and
$M(1^{+-})/M(0^{++})=1.91 \pm 0.05 \pm 0.12$.
\end{abstract}

\centerline{PACS number(s): 11.15.Ha, 12.38.Gc, 
14.20.-c,14.40.-n}

\eject


\section{INTRODUCTION}
\label{sec:level1}

If \rm{QCD} is the correct fundamental 
theory of strong interactions,  
a lot of glueball states should exist.
These glueballs are gluonic bound states
formed by strong self-interactions of gluons, 
and characterized by various quantum numbers $J^{PC}$. 
Precise calculation of the glueball masses will distinguish
\rm{QCD} from other models and give important guide for 
experimentalists. Among various nonperturbative techniques,
lattice QCD seems to give more reliable estimates
since it is model-independent.

Within lattice \rm{QCD}, 
Monte Carlo simulation on an Euclidean lattice
is still
the most popular method, 
since in the Euclidean formulation, space-time symmetry is 
apparent, the particle spectrum is given by the correlation length
of Green's function, and one can use the simulation methods in statistical
mechanics. In fact, it can easily bridge
the strong coupling region where confinement is obvious,
and weak coupling region which is more physically relevant.
With increasing lattice sizes, higher statistics and  
more suitable techniques, it might evidently give  
reliable predication for the glueball masses.
For a review of the current status,
see ref. \cite{Wein}.

The Hamiltonian formulation 
is an alternative to the standard Euclidean approach.
In such a formulation, one works with a conventional 
quantum mechanical system, and the spectrum and wavefunction
correspond to
an eigenvalue and eigenstate of the Hamiltonian.
Furthermore, it might possibly
give new insight into the structure of the theory.
Over the last twenty years, many attempts have been made.
The main difficulty of some conventional 
methods (e.g. strong coupling expansion) is that
they converge very slowly and 
very higher order $1/g^2$ calculations are
required to extend the results to the intermediate coupling region.
Unfortunately, high order calculations are difficult in practice.
For some other recent and interesting developments, see \cite{Hamer,SCHU} 
for details.

In refs. \cite{GCL,QCD3,MASS}, we proposed an efficient eigenvalue 
equation method for
Hamiltonian lattice gauge theory.  This method
consists of solving order by order the eigenvalue equation
with a suitable scheme preserving the correct continuum behavior
at any order of approximation.
So far, we have applied
this method to \rm{U(1)}, 
\rm{SU(2)} and \rm{SU(3)} models in 2+1 dimensions
and nonlinear $O(N)$ $\sigma$ model in 2 dimensions.
The reliability of our method is supported by increasing
evidences \cite{GCL,QCD3,MASS,CGZF,FLG,GCFC,SIGMA,COM}.
Most of these results have been summarized in \cite{Guo,Luo}.

In a recent paper \cite{LC}, we obtained the lightest glueball mass
in $\rm{QCD}_3$, more accurate than the previous papers \cite{MASS,COM}
with finite order truncation error under well control.  
We then used the idea of dimensional reduction 
\cite{Green,Samuel2} to estimate
the $0^{++}$ glueball mass in four dimensional QCD.
Our results suggested $M(0^{++}) \approx 1.71 \pm 0.05$ Gev, in
nice agreement with the most recent Monte Carlo data 
$M(0^{++}) \approx 1.740 \pm 0.071$ Gev by the IBM group \cite{IBM}.
This favors $\theta /f_J(1710)$
as a candidate of the $0^{++}$ glueball.

It is very desirable to do concrete computations 
in the realistic theory: \rm{QCD} in 3+1 dimensions.
In this paper, we try to make a first step towards this direction.
The rest of the paper is organized as follows.
In Sec.\ \ref{sec:level2},
we recapitulate the basic elements of our method.
The results for the glueball masses
are given in Sec.\ \ref{sec:level3}.
In the Appendix.\ \ref{sec:app}, we discuss the relation between
the Hamiltonian and Euclidean formulations for the purpose
of comparison.


\section{OUR METHOD}
\label{sec:level2}

Our method has been described in detail in refs. 
\cite{GCL,QCD3,MASS,CGZF,FLG,GCFC,SIGMA,COM,Guo,Luo}.
For the reader's convenience, we will repeat the most important
points.
The Schr{\"o}dinger
equation $H \vert \Omega \rangle 
= \epsilon_{\Omega} \vert \Omega \rangle$
on the Hamiltonian lattice
for the ground state 
\begin{eqnarray}
  \vert \Omega \rangle = exp \lbrack R(U) \rbrack \vert 0 \rangle
\label{b1}
\end{eqnarray}
and vacuum energy $\epsilon_{\Omega}$ can be reformulated as
\begin{eqnarray*}
\sum_{l} \lbrace [E_l,[E_l,R(U)]]+[E_l,R(U)][E_l,R(U)] \rbrace
\end{eqnarray*}
\begin{eqnarray}
- {2 \over g^4} \sum_{p} Tr(U_p+U_{p}^{\dagger})
={2a \over g^2} \epsilon_{\Omega}.
\label{schr}
\end{eqnarray}
To solve this equation, let us express $R(U)$ 
in order of independent set of
gauge invariant operators $G_{n,i}$, i.e.,
$  R(U)=\sum_{n} R_{n}(U)=\sum_{n,i} C_{n,i} G_{n,i}(U)$.
Substituting it to (\ref{schr}), 
we have the $N$th order truncated eigenvalue equation \cite{GCL,QCD3}
\begin{eqnarray*}
\sum_{l} \lbrace [E_l,[E_l,\sum_{n}^{N} R_{n}(U)]]
+\sum_{n_1+n_2 \le N}[E_l,R_{n_1}(U)][E_l,R_{n_2}(U)] \rbrace
\end{eqnarray*}
\begin{eqnarray}
- {4 \over g^4} \sum_{p} Re Tr(U_p)
={2a \over g^2} \epsilon_{\Omega}.
\label{b2}
\end{eqnarray}
Since the higher order graphs are created by the second term in (\ref{b2}), 
at $Nth$ order approximation all contributions from this term with 
$n_1+n_2>N$ should be discarded.
The reason for such a truncation method is to maintain the
correct continuum behavior during the truncation, as explained 
in detail in ref. \cite{GCL}.
By taking the coefficients 
of the graphs $G_{n,i}$ in this equation to zero,
we obtain a set of non-linear algebraic equations, from which
$C_{n,i}$ are determined.
Therefore, solving lattice field theory is reduced to solving the
algebraic equations.
Similarly, the eigenvalue equation 
$H \vert F \rangle = \epsilon_{F} \vert F \rangle$
for the glueball mass $M(J^{PC})=\Delta \epsilon
=\epsilon_F-\epsilon_{\Omega}$
and its wave function \cite{MASS} 
\begin{eqnarray}
\vert F \rangle = \lbrack F(U) - 
{\langle \Omega \vert F(U) \vert \Omega \rangle
\over \langle \Omega \vert \Omega \rangle} \rbrack \vert \Omega \rangle 
\label{WAVE}
\end{eqnarray}
at order $N$ is
\begin{eqnarray*}
\sum_{l} \lbrace [E_l,[E_l,\sum_{n}^{N} F_{n}(U)]]
+2\sum_{n_1+n_2 \le N}[E_l,F_{n_1}(U)][E_l,R_{n_2}(U)] \rbrace
\end{eqnarray*}
\begin{eqnarray}
={2a\Delta \epsilon \over g^2}  \sum_{n}^{N} F_{n}(U),
\label{c1}
\end{eqnarray}
from which the
coefficients $f_{n,i}$ in the glueball operator 
$F(U)=\sum_n F_{n}(U)=\sum_{n,i} f_{n,i} G^{J^{PC}}_{n,i}$ 
are determined. 
According to our experiences,
it is hopeful that the results converge rapidly to some stable value.
The difference between the results from different truncation orders 
is our estimate for the systematic error. 

Other detailed techniques, like 
the choice of independent set of operators in $SU(3)$ and
efficient ways to improve the continuum behavior 
of the physical quantities,
can be found in refs. \cite{QCD3,MASS,COM}. 


\section{RESULTS AND DISCUSSION}
\label{sec:level3}

Since this work is our first attempt to 
apply the method above to 3+1 dimensional pure QCD,
we would like to compute the masses of glueballs with
gluonic operators easily constructed. 
Here we choose three glueballs $0^{++}$, $0^{--}$, and $1^{+-}$.
Some lowest order gluonic operators are plotted in Fig. 1.
The operators $R_n$ and $F_n$ are linear combinations of these graphs,
and the first order ones are
\begin{eqnarray*}
R_1=C_1(G_{1,1}^{(1)}+G_{1,1}^{(2)}+G_{1,1}^{(3)})+h.c., 
\end{eqnarray*}
\begin{eqnarray}
F_{1}^{1^{+-}}=f_{1}^{1^{+-}}(G_{1,1}^{(3)}-h.c.).
\end{eqnarray}
Since the $0^{++}$ glueball has symmetry the same as the vacuum,
we take $F_{1}^{0^{++}}=R_1$. Once the lowest order graphs are determined,
the higher order graphs are generated by the second term in (\ref{b2})
for $R_n$ or that in (\ref{c1}) for $F_n$, from which an independent
set is chosen, taking into account the uni-modular conditions.
The operators so generated are certainly of the symmetry of the
corresponding glueballs. For example, at the $2nd$ order
\begin{eqnarray*}
F_2^{0^{++}}=f_{2,1}^{0^{++}}(G_{2,1}^{(1)}+G_{2,1}^{(2)}+G_{2,1}^{(3)})
+f_{2,2}^{0^{++}}(G_{2,2}^{(1)}+G_{2,2}^{(2)}+G_{2,2}^{(3)})
\end{eqnarray*}
\begin{eqnarray*}
+f_{2,3}^{0^{++}}(G_{2,3}^{(1)}+G_{2,3}^{(2)}+G_{2,3}^{(3)})
\end{eqnarray*}
\begin{eqnarray*}
+f_{2,4}^{0^{++}}(G_{2,4}^{(1)}+G_{2,4}^{(2)}+G_{2,4}^{(3)})
+f_{2,5}^{0^{++}}(G_{2,5}^{(1)}+G_{2,5}^{(2)}+G_{2,5}^{(3)})
\end{eqnarray*}
\begin{eqnarray*}
+f_{2,6}^{0^{++}}(G_{2,6}^{(1)}+G_{2,6}^{(2)}+G_{2,6}^{(3)})
\end{eqnarray*}
\begin{eqnarray*}
+f_{2,7}^{0^{++}}
(G_{2,7}^{(1)}+G_{2,7}^{(2)}+G_{2,7}^{(3)}+G_{2,7}^{(4)}
+G_{2,7}^{(5)}+G_{2,7}^{(6)}+G_{2,7}^{(7)}
+G_{2,7}^{(8)}+G_{2,7}^{(9)}
\end{eqnarray*}
\begin{eqnarray*}
+G_{2,7}^{(10)}
+G_{2,7}^{(11)}+G_{2,7}^{(12)})
\end{eqnarray*}
\begin{eqnarray*}
+f_{2,8}^{0^{++}}
(G_{2,8}^{(1)}+G_{2,8}^{(2)}+G_{2,8}^{(3)}+G_{2,8}^{(4)}
+G_{2,8}^{(5)}+G_{2,8}^{(6)}
+G_{2,8}^{(7)}+G_{2,8}^{(8)}+G_{2,8}^{(9)}
\end{eqnarray*}
\begin{eqnarray*}
+G_{2,8}^{(10)}
+G_{2,8}^{(11)}+G_{2,8}^{(12)})
\end{eqnarray*}
\begin{eqnarray*}
+f_{2,9}^{0^{++}}
(G_{2,9}^{(1)}+G_{2,9}^{(2)}+G_{2,9}^{(3)}+G_{2,9}^{(4)}
+G_{2,9}^{(5)}+G_{2,9}^{(6)}
+G_{2,9}^{(7)}
+G_{2,9}^{(8)}+G_{2,9}^{(9)}+G_{2,9}^{(10)}
\end{eqnarray*}
\begin{eqnarray*}
+G_{2,9}^{(11)}+G_{2,9}^{(12)})
\end{eqnarray*}
\begin{eqnarray*}
+f_{2,10}^{0^{++}}
(G_{2,10}^{(1)}+G_{2,10}^{(2)}+G_{2,10}^{(3)}+G_{2,10}^{(4)}
+G_{2,10}^{(5)}+G_{2,10}^{(6)}
\end{eqnarray*}
\begin{eqnarray*}
+G_{2,10}^{(7)}
+G_{2,10}^{(8)}+G_{2,10}^{(9)}+G_{2,10}^{(10)}
+G_{2,10}^{(11)}+G_{2,10}^{(12)})
+h.c.
\end{eqnarray*}
\begin{eqnarray*}
F_{2}^{0^{--}}=f_{2,7}^{0^{--}}
(G_{2,7}^{(1)}+G_{2,7}^{(2)}+G_{2,7}^{(3)}+G_{2,7}^{(4)}
+G_{2,7}^{(5)}+G_{2,7}^{(6)}
\end{eqnarray*}
\begin{eqnarray*}
+G_{2,7}^{(7)}
+G_{2,7}^{(8)}+G_{2,7}^{(9)}+G_{2,7}^{(10)}
+G_{2,7}^{(11)}+G_{2,7}^{(12)})
\end{eqnarray*}
\begin{eqnarray*}
+f_{2,8}^{0^{--}}(G_{2,8}^{(1)}+G_{2,8}^{(2)}+G_{2,8}^{(3)}+G_{2,8}^{(4)}
+G_{2,8}^{(5)}+G_{2,8}^{(6)}
+G_{2,8}^{(7)}
+G_{2,8}^{(8)}+G_{2,8}^{(9)}
\end{eqnarray*}
\begin{eqnarray*}
+G_{2,8}^{(10)}
+G_{2,8}^{(11)}+G_{2,8}^{(12)})
\end{eqnarray*}
\begin{eqnarray*}
+f_{2,9}^{0^{--}}(G_{2,9}^{(1)}
+G_{2,9}^{(2)}+G_{2,9}^{(3)}+G_{2,9}^{(4)}
+G_{2,9}^{(5)}+G_{2,9}^{(6)}+G_{2,9}^{(7)}
+G_{2,9}^{(8)}+G_{2,9}^{(9)}
\end{eqnarray*}
\begin{eqnarray*}
+G_{2,9}^{(10)}
+G_{2,9}^{(11)}+G_{2,9}^{(12)})
\end{eqnarray*}
\begin{eqnarray*}
+f_{2,10}^{0^{--}}(G_{2,10}^{(1)}
+G_{2,10}^{(2)}+G_{2,10}^{(3)}+G_{2,10}^{(4)}
+G_{2,10}^{(5)}+G_{2,10}^{(6)}
\end{eqnarray*}
\begin{eqnarray*}
+G_{2,10}^{(7)}
+G_{2,10}^{(8)}+G_{2,10}^{(9)}+G_{2,10}^{(10)}
+G_{2,10}^{(11)}+G_{2,10}^{(12)})
-h.c.
\end{eqnarray*}

\begin{eqnarray*}
F_{2}^{1^{+-}}=f_{2,1}^{1^{+-}}G_{2,1}^{(3)}
+f_{2,3}^{1^{+-}}G_{2,3}^{(3)}
+f_{2,4}^{1^{+-}}G_{2,4}^{(3)}
\end{eqnarray*}
\begin{eqnarray*}
+f_{2,7}^{1^{+-}}(G_{2,7}^{(5)}+G_{2,7}^{(6)}+G_{2,7}^{(7)}
+G_{2,7}^{(8)}+G_{2,7}^{(9)}+G_{2,7}^{(10)}
+G_{2,7}^{(11)}+G_{2,7}^{(12)})
\end{eqnarray*}
\begin{eqnarray*}
+f_{2,8}^{1^{+-}}
(G_{2,8}^{(5)}+G_{2,8}^{(6)}+G_{2,8}^{(7)}
+G_{2,8}^{(8)}+G_{2,8}^{(9)}+G_{2,8}^{(10)}
+G_{2,8}^{(11)}+G_{2,8}^{(12)})
\end{eqnarray*}
\begin{eqnarray*}
+f_{2,9}^{1^{+-}}
(G_{2,9}^{(5)}+G_{2,9}^{(6)}+G_{2,9}^{(7)}
+G_{2,9}^{(8)}+G_{2,9}^{(9)}+G_{2,9}^{(10)}
+G_{2,9}^{(11)}+G_{2,9}^{(12)})
\end{eqnarray*}
\begin{eqnarray*}
+f_{2,10}^{1^{+-}}
(G_{2,10}^{(5)}+G_{2,10}^{(6)}+G_{2,10}^{(7)}
+G_{2,10}^{(8)}+G_{2,10}^{(9)}+G_{2,10}^{(10)}
+G_{2,10}^{(11)}+G_{2,10}^{(12)})
\end{eqnarray*}
\begin{eqnarray}
-h.c.
\end{eqnarray}
Higher order graphs can be calculated in a similar way.

At relatively strong coupling, the absolute value of $aM_E$, 
calculated in the Hamiltonian formulation and converted to
the Euclidean one using 
(\ref{relation}), differs
from the result from strong coupling expansion 
in the Euclidean formulation. This is not surprising because
the Hamiltonian and Euclidean formulations are different schemes 
at finite lattice spacing,
and the weak coupling relation (\ref{relation}) doesn't not hold
in the strong coupling region. 
For $\beta \ge 6.0$, we do observe that the difference in $aM_E$ 
becomes much smaller at weaker coupling and there is a tendency 
approaching the Monte Carlo data \cite{Wein}. 

Similar to the most recent Monte Carlo data \cite{Wein} 
in the available coupling region,
clear asymptotic scaling window for the individual mass 
$aM(0^{++})$, $aM(0^{--})$, or $aM(1^{+-})$ could not be found. 
There might be two possible reasons for this scaling violation:
the available coupling region being not weak enough for 
the asymptotic scaling relation between $a$ and $\beta_E$ 
to be valid or the results being not accurate enough.
The first one may be reduced by the Symazik's improvement or
Lepage-Mackenzie scheme. The second one
may also be improved by higher order calculations.
For the mass ratio $M_E^{(1)}/M_E^{(2)}$, 
the large part of errors in $M_E^{(1)}$ and $M_E^{(2)}$ are
canceled.  Indeed, we observe approximate plateau for the mass ratio 
both for $M(0^{--})/M(0^{++})$ and $M(1^{+-})/M(0^{++})$
in the coupling region $\beta \in [6.0,6.4]$,
which are shown in Fig. 2. 

From the plateaus in $\beta \in [6.0,6.4]$,
we estimate 
\begin{eqnarray*}
{M(0^{--}) \over M(0^{++})}=2.44 \pm 0.05 \pm 0.20
\end{eqnarray*}
\begin{eqnarray}
{M(1^{+-}) \over M(0^{++})}=1.91 \pm 0.05 \pm 0.12,
\label{data}
\end{eqnarray}
where 
the mean value is the averaged one over the data in this region,
the first error is the error of the data in the plateau, and
the second error is the upper limit of the systematic errors
due to the finite order truncation.

For comparison, we list corresponding
results from other lattice calculations:
For $M(0^{--})/M(0^{++})$, it is approximately 
$3.43 \pm 1.50$ from Monte Carlo simulations (see the
references in \cite{Wein}), $2.2 \pm 0.2$ from
t-expansion plus Pad\'e approximants \cite{Horn}.
For $M(1^{+-})/M(0^{++})$, it is $1.88 \pm 0.02$ from Monte Carlo,
and $1.60 \pm 0.60$ from t-expansion plus Pad\'e.
For $M(0^{--})/M(0^{++})$, we obtain a value between the 
Monte Carlo and t-expansion data, with error under better control than
the former one. For the $M(1^{+-})/M(0^{++})$, where the systematic
error in the Monte Carlo data is very small, we get a value 
consistent well with the Monte Carlo data.

In conclusion, we have done spectrum computations 
of the $0^{++}$, $0^{--}$ and $1^{+-}$ glueballs, 
by means of the eigenvalue equation method in
\cite{GCL,QCD3,MASS}. 
Our first results for 3+1 dimensional QCD 
from this method are encouraging.
The spectrum calculations of other glueballs, such as 
$2^{++}$, $0^{-+}$ and $0^{+-}$,
are in progress.

\vskip 2cm

\noindent
{\bf Acknowledgments}

We thank T. Huang, J.M. Liu, C. Morningstar, D. Sch\"utte, X.Y. Shen,
J.M. Wu and some members of the BES collaboration
for useful communications.
This work was supported  
by the project of National Natural Science Foundation,
and that of
Chinese Education Committee. 

\vfill

\eject

\appendix
\section{EUCLIDEAN AND HAMILTONIAN FORMULATIONS}
\label{sec:app}


The standard Yang-Mills action is
\begin{eqnarray}
S=-{1 \over 4} \int d^4 x
\cal{F^{\alpha}_{\mu \nu}} \cal{F^{\alpha}_{\mu \nu}}.
\label{YM}
\end{eqnarray}
On the Euclidean lattice, the standard Wilson action is
\begin{eqnarray}
S=-{\beta_E \over 2N_c} \sum_p Tr(U_p+U_p^{\dagger}-2),
\label{Wilson}
\end{eqnarray}
where $\beta_E$ is related to the number of color $N_c$ and the bare 
coupling $g_E$ by $\beta_E=2N_c/g_E^2$, and $U_p \equiv U_{\mu \nu}$ 
is the product of four link variables around an
elementary plaquette in the $\mu$ and $\nu$ directions. 
In the continuum limit  
$\beta_E \to \infty$, it 
reduces to the Yang-Mills action (\ref{YM}).

To derive the lattice Hamiltonian from (\ref{Wilson}), 
let us fix the lattice spacing $a_s=a$ in spatial directions  
and decrease the lattice spacing $a_t$ in the temporal direction 
so that $\xi \equiv a/a_t \to \infty$.
In this case, (\ref{Wilson}) becomes  
\begin{eqnarray}
S=-{\beta_t \over 2N_c} \sum_{i} Tr(U_{it}+h.c.-2)-
{\beta_s \over 2N_c} \sum_{i>j} Tr(U_{ij}+h.c.-2).
\end{eqnarray}
Quantum-mechanically, 
\begin{eqnarray}
\beta_t={2N_c \over g_t^2}, ~~   \beta_s={2N_c \over g_s^2}
\end{eqnarray}
where $g_t \not= g_s$ due to the $\xi$ factor and quantum effects.
They are related to the Euclidean coupling by
\begin{eqnarray}
{1 \over g_t^2}={1 \over g_E^2}+c_t+O(g_E^2), ~~   
{1 \over g_s^2}={1 \over g_E^2}+c_s+O(g_E^2)
\end{eqnarray}
if $g_E$ is small enough.
The coefficients are calculated in \cite{Hasen}:
\begin{eqnarray}
c_t=4N_c[-0.01631+ {1 \over 32N_c^2}], ~~
c_s=4N_c[0.01707- {0.59173 \over 32N_c^2}].
\end{eqnarray}
Using the transfer matrix or Legendre transformation methods,
one can show that the lattice Hamiltonian in the temporal gauge is
\begin{eqnarray}
H={g_t \over g_s} H_{KS},
\end{eqnarray}
where
\begin{eqnarray}
H_{KS} = {g_H^{2} \over 2a} \sum_{x,j} 
E_{j}^{\alpha}(x) E_{j}^{\alpha}(x) 
- {1
\over ag_H^{2}} \sum_{p} Tr(U_{p} + U_{p}^{\dagger}-2).
\label{KS}
\end{eqnarray}
is the Kogut-Susskind Hamiltonian and
\begin{eqnarray} 
g_H= \sqrt{g_t g_s}
\label{gts}
\end{eqnarray}
is the gauge coupling 
of the Kogut-Susskind Hamiltonian gauge theory. 
In (\ref{KS}), the summation $\sum_{x,j}$ is over the link $l$ 
in the positive spatial directions, and the summation $\sum_p$ is
over the spatial elementary plaquette.
For weak enough coupling $g_E \to 0$, the mass can be compared with 
the Euclidean one
\begin{eqnarray} 
M_E=M_H {g_t \over g_s}
\label{relation}
\end{eqnarray}
if $g_H$ is converted to $g_E$ by (\ref{gts}). The $g_t/g_s$
factor doesn't matter if one calculates the mass ratio.

\vfill
\eject

\vfill


\centerline{\bf Figure Captions}

\vskip 3.0cm

\noindent
Fig.1 Some relevant gluonic operators at the lowest orders.

\vskip 2.0cm

\noindent
Fig. 2 Mass ratios of the glueballs.

\end{document}